# Two-dimensional electron-hole system under the influence of the Chern-Simons gauge field created by the quantum point vortices

S.A. Moskalenko[1], V.A. Moskalenko[1], I.V. Podlesny[1] and M.A. Liberman[2]

[1]*Institute of Applied Physics, 5, Academiei str., MD-2028 Chisinau, Republic of Moldova*
[2]*Nordic* Institute *for Theoretical Physics (NORDITA) KTH and Stockholm University, Roslagstullsbacken 23, Stockholm, SE-106 91 Sweden*

**Abstract**

In the present work the Chern-Simons(C-S) gauge field theory developed by Jackiw and Pi [1] and widely used to explain the fractional quantum Hall effects, was applied to describe the two-dimensional (2D) electron-hole (e-h) system in a strong perpendicular magnetic field under the influence of the quantum point vortices creating the Chern-Simons(C-S) gauge field. The composite particles formed by electrons and by holes with equal integer positive numbers $\phi$ of the attached quantum point vortices are described by the dressed field operators, which obey to the Fermi or to the Bose statistics depending on the even or odd numbers $\phi$. It is shown that the phase operators as well as the vector and the scalar potentials of the C-S gauge field depend on the difference of the electron and of the hole density operators. They vanish in the mean field approximation, when the average values of the electron and of the whole densities coincide. Nevertheless, even in this case, the quantum fluctuations of the C-S gauge field lead to new physics of the 2D e-h system.

## § 1. Introduction

In Ref [1] Jackiw and Pi developed the classical and the quantum nonrelativistic Chern-Simons (C-S) theory for the two-dimensional N-body system of point particles. The Chern-Simons gauge field is created by the quantum point vortices localized at each charged particle. In Jackiw and Pi [1] described the dynamics of N-particles moving in the plane and interacting through the mediation of the $u(1)$ gauge field with C-S singular phase and with kinetic action. In Refs. [2-4] the C-S theory was used to describe the fractional quantum Hall effects. The phase operator $\hat{\omega}(\vec{r})$ of the C-S field was introduced as a coherent summation of the angles $\theta(\vec{r}-\vec{r}')$ formed with the in-plane x-axis by the reference vectors $(\vec{r}-\vec{r}')$, which determine positions $\vec{r}'$ of the particles, creating the gauge field in the point $\vec{r}$. It was pointed in Ref [1] that angles $\theta(\vec{r}-\vec{r}')$ are ill determined because arctangent is a multivalued function. But this deficiency was compensated by the fact, that the summation of the angles was weighted in Ref. [1] by the density operators $\hat{\rho}(\vec{r}')$ of the charged particles in the way

$$\hat{\omega}(\vec{r}) = -\frac{\phi e}{\alpha}\int d\vec{r}'\, \theta(\vec{r}-\vec{r}')\hat{\rho}(\vec{r}'); \quad \theta(\vec{r}-\vec{r}') = \arctan\left(\frac{y-y'}{x-x'}\right). \tag{1}$$

Here $\phi$ is an integer, positive number and $\alpha$ is the fine structure constant $\alpha = e^2/\hbar c = 1/137$. The integer value of $\phi$ is another factor, which makes the dressed field operators to be well defined. Because we are interested in the generalization of the C-S theory from the one component electron gas to the two-component e-h system, we repeat the main statements of the C-S theory in a new variant, introducing from the very beginning a supplementary label $i = e, h$ denoting electrons and holes. The partial field operators $\Psi_i(r)$ and $\Psi_i^+(r)$ lead to the partial phase operators $\hat{\omega}_i(\vec{r})$ and to the partial vector potential operators $\hat{\vec{a}}_i(\vec{r})$ taking into account the electrical charges of the electrons (-e) and of the holes (+e).

The bare field operators will be denoted as $\hat{\Psi}_i^0(\vec{r})$, $\Psi_i^{0+}(\vec{r})$ with supplementary label zero, whereas the dressed field operator $\hat{\Psi}_i(\vec{r})$ and $\Psi_i^+(\vec{r})$ will be written without it. Note that while the bare and the dressed field operators are different their density operators $\hat{\rho}_i(\vec{r})$ and $\hat{\rho}_i^0(\vec{r})$ coincide

$$\hat{\rho}_i(\vec{r}) = \hat{\Psi}_i^+(\vec{r})\hat{\Psi}_i(\vec{r}) = \hat{\Psi}_i^{+0}(\vec{r})\hat{\Psi}_i^0(\vec{r}) = \hat{\rho}_i^0(\vec{r}) \tag{2}$$

Below we will show that this property is the consequence of the unitary transformation $\hat{u}^+(r)\hat{u}(r)=1$ and it concerns any operators, which are analytical functions of the density operators in the way $f(\hat{\rho}_i(\vec{r})) = f(\hat{\rho}_i^0(\vec{r}))$.

For example, the partial phase operators $\hat{\omega}_i(\vec{r})$ and the partial vector potential operators $\hat{\vec{a}}_i(\vec{r})$ demonstrate this property

$$\hat{\omega}_i(\vec{r}) = -\frac{\phi e}{\alpha}\int d^2\vec{r}'\,\theta(\vec{r}-\vec{r}')\hat{\rho}_i(\vec{r}') = \hat{\omega}_i^0(\vec{r}),$$

$$\hat{\vec{a}}_i(\vec{r}) = \vec{\nabla}_{\vec{r}}\hat{\omega}_i(\vec{r}) = -\frac{\phi e}{\alpha}\int d^2\vec{r}'\,\vec{\nabla}_{\vec{r}}\theta(\vec{r}-\vec{r}')\hat{\rho}_i(\vec{r}') = \hat{\vec{a}}_i^0(\vec{r}), \quad (3)$$

$$\hat{\omega}(\vec{r}) = \hat{\omega}_e(\vec{r}) - \hat{\omega}_h(\vec{r}),\ \hat{\vec{a}}(\vec{r}) = \hat{\vec{a}}_e(\vec{r}) - \hat{\vec{a}}_h(\vec{r}).$$

The differences added in (3) give rise to the resultant phase and vector potential operators created by the integer e-h system.

The phase operators $\hat{\omega}_i(\vec{r})$ and $\hat{\omega}(\vec{r})$ are singular values because they are expressed through the multivalued function such as arctangent and therefore as was pointed in Ref [1] the function $\theta(\vec{r}-\vec{r}')$ is ill determined. Some of its properties are:

$$\vec{\nabla}_{\vec{r}}\theta(\vec{r}-\vec{r}') = -\vec{\nabla}_{\vec{r}}\times\ln|\vec{r}-\vec{r}'|;\ \vec{\nabla}\times\theta(\vec{r}-\vec{r}') = \vec{\nabla}\ln(\vec{r}-\vec{r}'),$$

$$\Delta_{\vec{r}}\theta(\vec{r}-\vec{r}') = 0,\ \Delta_{\vec{r}}\ln|\vec{r}-\vec{r}'| = 2\pi\delta^2(\vec{r}-\vec{r}');$$

where

$$\vec{\nabla}\times = \vec{e}_x\frac{\partial}{\partial y} - \vec{e}_y\frac{\partial}{\partial x};\ \vec{\nabla} = \vec{e}_x\frac{\partial}{\partial x} + \vec{e}_y\frac{\partial}{\partial y} = \frac{\vec{\in}_\theta}{r}\frac{\partial}{\partial \theta} + \vec{\in}_{\vec{r}}\frac{\partial}{\partial r};\ \vec{\in}_\theta = \frac{-\vec{e}_x y + \vec{e}_y x}{r};\ \vec{\in}_{\vec{r}} = \frac{\vec{r}}{r},$$

$$\vec{\nabla}\times\vec{V} = \in^{ij}\partial_i V_j = \frac{\partial}{\partial x}V_y - \frac{\partial}{\partial y}V_x = S,$$

$$(\vec{\nabla}\times S)^i = \in^{ij}\partial_j S;$$

$$\in^{12} = -\in^{21} = 1,$$

$$\in^{11} = \in^{22} = 0.$$

$$\vec{\nabla}\hat{\vec{a}}(\vec{r}) = 0,$$

$$\vec{\nabla}\times\hat{\vec{a}}(\vec{r}) = \hat{b}(\vec{r}) = \vec{\nabla}_{\vec{r}}\times\frac{\phi e}{\alpha}\int d^2\vec{r}'\vec{\nabla}_{\vec{r}}\times\ln|\vec{r}-\vec{r}'|\hat{\rho}(\vec{r}') =$$

$$= \frac{\phi e}{\alpha}\int d^2\vec{r}'\Delta_{\vec{r}}\ln|\vec{r}-\vec{r}'|\hat{\rho}(\vec{r}') = 2\pi\frac{\phi e}{\alpha}\hat{\rho}(\vec{r}), \quad (4)$$

$$\hat{b}(\vec{r}) = 2\pi\frac{\phi e}{\alpha}\hat{\rho}(\vec{r}) = 2\pi\frac{\phi e}{\alpha}(\hat{\rho}_e(\vec{r}) - \hat{\rho}_h(\vec{r})).$$

It was pointed out in Ref [1] that, in the 2D space the curl of the vector is a scalar, whereas the curl of the scalar is a vector. These properties are demonstrated by formulas (4), in particular, the Green function of the Laplacian in the 2D space is $(\ln \vec{r})/2\pi$. Another important information shown by formulas (4) is the effective magnetic field $\hat{\vec{b}}(\vec{r})$ expressed by $\vec{\nabla} \times \hat{\vec{a}}(\vec{r})$. It was shown in [1] that this magnetic field is created by the quantum point vortices. In the case of the one-component electron gas this supplementary magnetic field may compensate the external magnetic field. As it will be shown below in the case of the two-component e-h system this effective magnetic field has a special interesting property. It seems to vanish in the mean-field approximation when the average values of the density operators coincide $\langle \hat{\rho}_e(\vec{r}) \rangle = \langle \hat{\rho}_h(\vec{r}) \rangle$, but its quantum fluctuations lead to unexpected physics of the 2D e-h system in the strong external magnetic field. Jackiw and Pi [1] accorded a special attention to the calculations with the participation of the ill determined angle function $\theta(\vec{r} - \vec{r}')$. They stressed that in the nonrelativistic quantum mechanics the particles are points and the density operator $\hat{\rho}(\vec{r})$ is localized in these points being a superposition of the $\delta$ functions. This fact plays a critical role in the calculations involving the C-S gauge field. For example, it permits to interchange the integration and the differentiation in the definition of the C-S vector potential $\hat{\vec{a}}(\vec{r})$. Otherwise it would be impossible to move the gradient with respect to $\vec{r}$ out of the integral on the variable $\vec{r}'$. In general case the operators (3) are singular because the angle $\theta(\vec{r} - \vec{r}')$ is a multivalued function and the integration over 2D $\vec{r}'$ plane requires specifying of the cut in $\vec{r}'$ beginning in $\vec{r}$. However the presence of the density operator $\hat{\rho}(\vec{r}')$ in the integral with $\delta$-function eigenvalues leads to the exceptional situation, when $\vec{r}$ -gradient can be moved with impunity outside the integral. Because the derivative of the function $\theta(\vec{r} - \vec{r}')$ in the point $\vec{r} = \vec{r}'$ is ill defined, in [1] was proposed its regularization $\vec{\nabla}_{\vec{r}} = \theta(\vec{r} - \vec{r}')\big|_{\vec{r}=\vec{r}'} = 0$.

To confirm the affirmation concerning the eigenvalues of the density operators one should use the commutation relations between the field operators and the density operator:

$$\left[\Psi_i(r), \hat{\rho}_i(\vec{r}')\right] = \delta^2(\vec{r} - \vec{r}')\hat{\Psi}_i(\vec{r}'); \quad \left[\Psi_i^+(r), \hat{\rho}_i(\vec{r}')\right] = -\delta^2(\vec{r} - \vec{r}')\hat{\Psi}_i^+(\vec{r}'),$$

$$\left[\hat{\rho}_i(\vec{r}), \hat{\rho}_j(\vec{r}')\right] = 0, \quad \left[\hat{\rho}_i(\vec{r}), \hat{\omega}(\vec{r}')\right] = 0, \quad \left[\hat{\omega}(\vec{r}'), \hat{\vec{a}}(\vec{r})\right] = 0. \tag{5}$$

The proper functions of the density operators $\hat{\rho}_i(\vec{r})$ can be introduced in the way

$$\left|\hat{\Psi}_i\left(\vec{r}'\right)\right\rangle = \hat{\Psi}_i^+\left(\vec{r}'\right)|0\rangle, \quad \left\langle\hat{\Psi}_i\left(\vec{r}'\right)\right| = \langle 0|\hat{\Psi}_i\left(\vec{r}'\right), \tag{6}$$

where $|0\rangle$ is the ground state of the system. The action of the density operator on the function $\left|\hat{\Psi}_i\left(\vec{r}'\right)\right\rangle$ gives rise to the result

$$\begin{aligned}\hat{\rho}_i(\vec{r})\left|\hat{\Psi}_i\left(\vec{r}'\right)\right\rangle &= \hat{\Psi}_i^+(\vec{r})\hat{\Psi}_i(\vec{r})\hat{\Psi}_i^+\left(\vec{r}'\right)|0\rangle = \\ &= \delta^2\left(\vec{r}-\vec{r}'\right)\hat{\Psi}_i^+(\vec{r})|0\rangle = \delta^2\left(\vec{r}-\vec{r}'\right)\left|\hat{\Psi}_i\left(\vec{r}'\right)\right\rangle.\end{aligned} \tag{7}$$

It confirms that the eigenvalues of the density operators $\hat{\rho}_i(\vec{r})$ have the forms of $\delta$-functions as well as the decisive role played by these operators to combat the deficiencies related with the presence of the angle function $\theta(\vec{r}-\vec{r}')$. The integer, positive values of the numbers $\phi$ in the definitions of the operators (3) contribute also to remove the incertitude related with the multivalued character of the angles $\theta(\vec{r}-\vec{r}')$.

These introductory details will be completed as follows. Section 2 describes the unitary transformation operators that introduce the Chern-Simons field into a two-component e-h-system. In Section 3 we derive the Hamiltonian and the equations of motion for the dressed field operators. The new properties of the 2D magnetoexcitons under the influence of the C-S gauge field are revealed in Section 4. We conclude in Section 5.

## §2. The unitary transformation introducing the Chern-Simons gauge field in the two-component electron-hole system

In the two-component e-h system the electrons and the holes contribute in an equal manner to the creation of the unique and a common C-S gauge field, each of them acting with the proper electric charge. The resultant phase operator $\hat{\omega}(\vec{r})$ as well as its gradient are algebraic sums of the partial electron and hole contributions

$$\begin{aligned}\hat{\omega}(\vec{r}) &= \hat{\omega}_e(\vec{r}) - \hat{\omega}_h(\vec{r}) = -\frac{\phi e}{\alpha}\int d^2\vec{r}'\,\theta(\vec{r}-\vec{r}')\hat{\rho}(\vec{r}') = \hat{\omega}^+(\vec{r}), \\ \hat{\vec{a}}(\vec{r}) &= \hat{\vec{a}}_e(\vec{r}) - \hat{\vec{a}}_h(\vec{r}) = -\frac{\phi e}{\alpha}\int d^2\vec{r}'\,\vec{\nabla}_{\vec{r}}\,\theta(\vec{r}-\vec{r}')\hat{\rho}(\vec{r}') = \hat{\vec{a}}^+(\vec{r}), \\ \hat{\rho}(\vec{r}) &= \hat{\rho}_e(\vec{r}) - \hat{\rho}_h(\vec{r}) = \hat{\rho}^+(\vec{r}); \\ \hat{\rho}_i(\vec{r}) &= \hat{\Psi}_i^+(\vec{r})\hat{\Psi}_i(\vec{r}) = \hat{\rho}_i^+(\vec{r}).\end{aligned} \tag{8}$$

These expressions are true in the bare as well as in the dressed representations no matter to which statistics the Fermi or to the Bose ones obey the field operators. Contrary to the one component two-dimensional electron gas (2DEG), in our case there are two subsystems with different electric charges. As the consequence, the resultant phase and the vector potential operators compensate each other, so as to obtain in the mean-field approximation a zero-gauge field. It opens the possibility to neglect in the zero-order approximation by effects arising due to the influence of the C-S gauge field, taking them into account in the next orders of the perturbation theory. In the case of the two-component electron-hole system, the unitary transformation introducing the C-S gauge field looks as follows:

$$\hat{u}(r) = e^{\frac{ie}{\hbar c}\hat{\omega}(\vec{r})}; \quad \hat{u}^+(r) = e^{-\frac{ie}{\hbar c}\hat{\omega}(\vec{r})}; \quad \hat{u}^+(r)\hat{u}(r) = 1. \tag{9}$$

The bare electron and hole field operators will be denoted as $\hat{\Psi}_i^0(\vec{r})$ and $\hat{\Psi}_i^{0+}(\vec{r})$ with the supplementary zero label. They obey to the Fermi statistics with the Fermi commutation relations.

$$\hat{\Psi}_i^0(\vec{r})\hat{\Psi}_j^{0+}(\vec{r}) + \hat{\Psi}_j^{0+}(\vec{r})\hat{\Psi}_i^0(\vec{r}) = \delta_{ij}\delta^2(\vec{r}-\vec{r}'),$$
$$\hat{\Psi}_i^0(\vec{r})\hat{\Psi}_j^0(\vec{r}') + \hat{\Psi}_j^0(\vec{r}')\hat{\Psi}_i^0(\vec{r}) = 0. \tag{10}$$

The dressed electron and hole field operators creating the C-S gauge field are written without the zero label and they are introduced in the form

$$\hat{\Psi}_e(\vec{r}) = u^+(\vec{r})\hat{\Psi}_e^0(\vec{r}), \quad \hat{\Psi}_e^+(\vec{r}) = \hat{\Psi}_e^{0+}(\vec{r})u(\vec{r}),$$

$$\hat{\Psi}_h(\vec{r}) = \hat{u}(\vec{r})\hat{\Psi}_h^0(\vec{r}), \quad \hat{\Psi}_h^+(\vec{r}) = \hat{\Psi}_h^{0+}(\vec{r})u^+(\vec{r}),$$

$$\hat{\rho}_i(\vec{r}) = \hat{\Psi}_i^+(\vec{r})\hat{\Psi}_i(\vec{r}) = \hat{\rho}_i^0(\vec{r}) = \hat{\Psi}_i^{0+}(\vec{r})\hat{\Psi}_i^0(\vec{r}), i = e, h;$$

$$\hat{\rho}_i^+(\vec{r}) = \hat{\rho}_i(\vec{r}), \quad \hat{\rho}_i^{0+}(\vec{r}) = \hat{\rho}_i^0(\vec{r}), \quad [\hat{\rho}_i(\vec{r}), \hat{\rho}_j(\vec{r}')] = 0, \tag{11}$$

$$\hat{\rho}(\vec{r}) = \hat{\rho}_e(\vec{r}) - \hat{\rho}_h(\vec{r}) = \hat{\rho}^0(\vec{r}) = \hat{\rho}_e^0(\vec{r}) - \hat{\rho}_h^0(\vec{r}).$$

Due to the equality $\hat{\rho}(\vec{r}) = \hat{\rho}^0(\vec{r})$, the phase and the vector potential operators determined by formulas (8) are the same in the bare and in the dressed representations:

$$\hat{\omega}(\vec{r}) = \hat{\omega}^0(\vec{r}), \quad \hat{\omega}_i(\vec{r}) = \hat{\omega}_i^0(\vec{r}),$$
$$\hat{\vec{a}}(\vec{r}) = \hat{\vec{a}}^0(\vec{r}), \quad \hat{\vec{a}}_i(\vec{r}) = \hat{\vec{a}}_i^0(\vec{r}), \quad \hat{\rho}(\vec{r}) = \hat{\rho}^0(\vec{r}). \tag{12}$$

This concerns also to other operators expressed as the analytic functions depending on the density operator $\hat{\rho}(\vec{r})$. In spite of the fact that the bare electron and hole field operators $\hat{\Psi}_i^0(\vec{r})$ and $\hat{\Psi}_i^{0+}(\vec{r})$ obey to the Fermi commutation relations (10), the dressed field operators $\hat{\Psi}_i(\vec{r})$ and $\hat{\Psi}_i^+(\vec{r})$ satisfy to the Fermi or to the Bose statistics depending on the even or the odd integer, positive numbers $\phi$ introduced into the definitions of operators (8). The proving of this statement needs to derive first the commutation relations between the field operators $\hat{\Psi}_i(\vec{r})$ and $\hat{\Psi}_i^+(\vec{r})$ with the density operators $\hat{\rho}_i(\vec{r})$, as well as the commutation relations between the field operators and the unitary transformations operators $u(\vec{r})$ and $u^+(\vec{r})$. The first of them are as follows

$$\left[\hat{\Psi}_i(\vec{r}), \hat{\rho}_i(\vec{r}')\right] = \delta^2(\vec{r}-\vec{r}')\hat{\Psi}_i(\vec{r}'); \quad \left[\hat{\Psi}_i^+(\vec{r}'), \hat{\rho}_i(\vec{r}')\right] = -\delta^2(\vec{r}-\vec{r}')\hat{\Psi}_i^+(\vec{r}'). \tag{13}$$

They are the same as in the case of the Fermi or of the Bose statistics. On this base one can obtain the following commutation relations

$$\hat{\Psi}_e(\vec{r})\hat{\omega}^n(\vec{r}') = \left[\hat{\omega}(\vec{r}') - \frac{\phi e}{\alpha}\theta(\vec{r}'-\vec{r})\right]^n \hat{\Psi}_e(\vec{r}),$$

$$\hat{\Psi}_h(\vec{r})\hat{\omega}^n(\vec{r}') = \left[\hat{\omega}(\vec{r}') + \frac{\phi e}{\alpha}\theta(\vec{r}'-\vec{r})\right]^n \hat{\Psi}_h(\vec{r}),$$

$$\hat{\Psi}_e^+(\vec{r})\hat{\omega}^n(\vec{r}') = \left[\hat{\omega}(\vec{r}') + \frac{\phi e}{\alpha}\theta(\vec{r}'-\vec{r})\right]^n \hat{\Psi}_e^+(\vec{r}),$$

$$\hat{\Psi}_h^+(\vec{r})\hat{\omega}^n(\vec{r}') = \left[\hat{\omega}(\vec{r}') - \frac{\phi e}{\alpha}\theta(\vec{r}'-\vec{r})\right]^n \hat{\Psi}_h^+(\vec{r}). \tag{14}$$

Now the commutation relations of the field operators $\hat{\Psi}_i(\vec{r})$ with the unitary transformation operators $e^{\pm\frac{ie}{\hbar c}\hat{\omega}(\vec{r}')}$ will be

$$\hat{\Psi}_e(\vec{r})e^{\pm\frac{ie}{\hbar c}\hat{\omega}(\vec{r}')} = \sum_{n=0}^{\infty}\left(\frac{\pm ie}{\hbar c}\right)^n \frac{1}{n!}\hat{\Psi}_e(\vec{r})\hat{\omega}^n(\vec{r}') =$$

$$= \sum_{n=0}^{\infty}\left(\frac{\pm ie}{\hbar c}\right)^n \frac{1}{n!}\left[\hat{\omega}(\vec{r}') - \frac{\phi e}{\alpha}\theta(\vec{r}'-\vec{r})\right]^n \hat{\Psi}_e(\vec{r}) =$$

$$= e^{\mp i\phi\theta(\vec{r}'-\vec{r})}e^{\pm\frac{ie}{\hbar c}\hat{\omega}(\vec{r}')}\hat{\Psi}_e(\vec{r}),$$

$$\hat{\Psi}_h(\vec{r})e^{\pm\frac{ie}{\hbar c}\hat{\omega}(\vec{r}')} = e^{\pm i\phi\theta(\vec{r}'-\vec{r})}e^{\pm\frac{ie}{\hbar c}\hat{\omega}(\vec{r}')}\hat{\Psi}_h(\vec{r}),$$

$$\hat{\Psi}_e^+(\vec{r})e^{\pm\frac{ie}{\hbar c}\hat{\omega}(\vec{r}')} = e^{\pm i\phi\theta(\vec{r}'-\vec{r})}e^{\pm\frac{ie}{\hbar c}\hat{\omega}(\vec{r}')}\hat{\Psi}_e^+(\vec{r}), \tag{15}$$

$$\hat{\Psi}_h^+(\vec{r})e^{\pm\frac{ie}{\hbar c}\hat{\omega}(\vec{r}')} = e^{\mp i\phi\theta(\vec{r}'-\vec{r})}e^{\pm\frac{ie}{\hbar c}\hat{\omega}(\vec{r}')}\hat{\Psi}_h^+(\vec{r}).$$

To prove the main statement concerning the statistics of the dressed field operators we will start with the first equation (10) and will transcribe it from the bare to the dressed operators in the way

$$\hat{\Psi}_e^0(\vec{r})\hat{\Psi}_e^{0+}(\vec{r}') + \hat{\Psi}_e^{0+}(\vec{r}')\hat{\Psi}_e^0(\vec{r}) = \delta^2(\vec{r}-\vec{r}') =$$
$$= u(\vec{r})\hat{\Psi}_e(\vec{r})\hat{\Psi}_e^+(\vec{r}')u^+(\vec{r}') + \hat{\Psi}_e^+(\vec{r}')u^+(\vec{r}')u(\vec{r})\hat{\Psi}_e(\vec{r}) =$$
$$= \hat{\Psi}_e(\vec{r})\hat{\Psi}_e^+(\vec{r}')u(\vec{r})u^+(\vec{r}')e^{i\phi\theta(0)}e^{-i\phi\theta(\vec{r}-\vec{r}')} +$$
$$+ \hat{\Psi}_e^+(\vec{r}')\hat{\Psi}_e(\vec{r})u^+(\vec{r}')u(\vec{r})e^{i\phi\theta(0)}e^{-i\phi\theta(\vec{r}'-\vec{r})},$$
$$u(\vec{r})u^+(\vec{r}') = u^+(\vec{r}')u(\vec{r}),$$
$$\theta(\vec{r}'-\vec{r}) = \theta(\vec{r}-\vec{r}') + \pi.$$
(16)

With account of the last two relations the equation (16) can be transcribed in the form

$$e^{i\phi\theta(0)}e^{-i\phi\theta(\vec{r}-\vec{r}')}\left[\hat{\Psi}_e(\vec{r})\hat{\Psi}_e^+(\vec{r}') + e^{-i\phi\pi}\hat{\Psi}_e^+(\vec{r}')\hat{\Psi}_e(\vec{r})\right] \times$$
$$\times u(\vec{r})u^+(\vec{r}') = \delta^2(\vec{r}-\vec{r}').$$
(17)

It is equivalent to the commutation relation

$$\hat{\Psi}_e(\vec{r})\hat{\Psi}_e^+(\vec{r}') + e^{-i\phi\pi}\hat{\Psi}_e^+(\vec{r}')\hat{\Psi}_e(\vec{r}) = \delta^2(\vec{r}-\vec{r}'),$$
$$e^{-i\phi\pi} = \cos\phi\pi - i\sin(\phi\pi) = \begin{cases} 1, \phi = 0, 2, 4..., F \\ -1, \phi = 1, 3, 5.., B \end{cases}.$$
(18)

The most important result of these calculation is the affirmation that the C-S gauge field operators $\hat{\Psi}_i^+(\vec{r})$ and $\hat{\Psi}_i(\vec{r})$ with $i = e, h$ obey the Fermi statistics in the case of the even integer, positive pair numbers $\phi$ and to the Bose statistics in the case of odd integer positive numbers $\phi$. It is an important result of the Chern-Simons gauge field theory developed by Jackiw and Pi in Ref [1].

### §3. The Hamiltonian and the equations of motion describing the dressed operators of the C-S gauge field.

To obtain the Hamiltonian describing the interaction of the composite particles expressed through the dressed field operators $\hat{\Psi}_i^+(\vec{r})$ and $\hat{\Psi}_i(\vec{r})$, as well as to deduce their equations of motion, we will start with the corresponding expressions for the bare field operators $\hat{\Psi}_i^{0+}(\vec{r})$ and $\hat{\Psi}_i^0(\vec{r})$. The Hamiltonian describing the bare 2D electrons and holes in the external perpendicular magnetic field and interacting by the Coulomb forces obtained in Ref. [9] looks as follows

$$\hat{H}^0 = \hat{K}^0 + \hat{H}^0_{Coul.},$$

$$\hat{K}^0 = \frac{\hbar^2}{2m_e}\int d^2\vec{r}\,'\hat{\Psi}_e^{0+}(\vec{r}\,')\left(-i\vec{\nabla}' + \frac{e}{\hbar c}\vec{A}(\vec{r}\,')\right)^2 \hat{\Psi}_e^{0+}(\vec{r}\,') +$$

$$+ \frac{\hbar^2}{2m_h}\int d^2\vec{r}\,'\hat{\Psi}_h^{0+}(\vec{r}\,')\left(-i\vec{\nabla}' - \frac{e}{\hbar c}\vec{A}(\vec{r}\,')\right)^2 \hat{\Psi}_h^{0+}(\vec{r}\,'),$$

$$\hat{H}^0_{Coul.} = \frac{1}{2}\int d^2\vec{r}\,'\int d^2\vec{r}\,''V_{Coul.}(\vec{r}\,' - \vec{r}\,'')\hat{\Psi}_e^{0+}(\vec{r}\,')\hat{\rho}_e^0(\vec{r}\,'')\hat{\Psi}_e^0(\vec{r}\,') +$$

$$+ \frac{1}{2}\int d^2\vec{r}\,'\int d^2\vec{r}\,''V_{Coul.}(\vec{r}\,' - \vec{r}\,'')\hat{\Psi}_h^{0+}(\vec{r}\,')\hat{\rho}_h^0(\vec{r}\,'')\hat{\Psi}_h^0(\vec{r}\,') -$$

$$- \int d^2\vec{r}\,'\int d^2\vec{r}\,''V_{Coul.}(\vec{r}\,' - \vec{r}\,'')\hat{\Psi}_e^{0+}(\vec{r}\,')\hat{\rho}_h^0(\vec{r}\,'')\hat{\Psi}_e^0(\vec{r}\,').$$

(19)

Here $\vec{A}(\vec{r}\,')$ is the vector potential created by the external magnetic field perpendicular to the layer. In the Landau gauge description, it has the form $\vec{A}(\vec{r}) = (-B \cdot y, 0, 0)$, where $B$ is the magnetic field strength. The vector potential obeys the condition $\vec{\nabla}\vec{A}(\vec{r}) = 0$. The Coulomb interaction potential in the 2D system can be presented as

$$V_{Coul.}(\vec{r}) = \sum_{\vec{Q}} V_{\vec{Q}} e^{i\vec{Q}\vec{r}}, \quad V(\vec{Q}) = \frac{2\pi e^2}{\varepsilon_0 S |\vec{Q}|}, \quad W(\vec{Q}) = V(\vec{Q}) e^{-\frac{Q^2 l_0^2}{2}}; \quad l_0^2 = \frac{\hbar c}{eB}. \quad (20)$$

Here $S$ is the layer surface area, $\varepsilon_0$ is the effective dielectric constant and $l_0$ is the magnetic length. Side by side with the coefficient $V(\vec{Q})$, in (20) was introduced also the coefficient $W(\vec{Q})$, which will be used below.

The bare density operators $\hat{\rho}_i^0(\vec{r})$ were introduced by the formula (2). The Schrodinger equations for the bare operators $\hat{\Psi}_e^0(\vec{r})$ and $\hat{\Psi}_h^0(\vec{r})$ were derived in the Ref [9], and we just recall them:

$$i\hbar \frac{d\hat{\Psi}_e^0(\vec{r})}{dt} = \left[\hat{\Psi}_e^0(\vec{r}), \hat{H}^0\right] = \frac{\hbar^2}{2m_e}\left(-i\vec{\nabla} + \frac{e}{\hbar c}\vec{A}(\vec{r})\right)^2 \hat{\Psi}_e^0(\vec{r}) +$$

$$+\int d^2\vec{r}' V_{Coul.}(\vec{r}-\vec{r}')\hat{\rho}^0(\vec{r}')\hat{\Psi}_e^0(\vec{r}),$$

$$i\hbar \frac{d\hat{\Psi}_h^0(\vec{r})}{dt} = \left[\hat{\Psi}_h^0(\vec{r}), \hat{H}^0\right] = \frac{\hbar^2}{2m_h}\left(-i\vec{\nabla} - \frac{e}{\hbar c}\vec{A}(\vec{r})\right)^2 \hat{\Psi}_h^0(\vec{r}) -$$

$$-\int d^2\vec{r}' V_{Coul.}(\vec{r}-\vec{r}')\hat{\rho}^0(\vec{r}')\hat{\Psi}_h^0(\vec{r}),$$

$$i\hbar \frac{d\hat{\Psi}_e^{0+}(\vec{r})}{dt} = \left[\hat{\Psi}_e^{0+}(\vec{r}), \hat{H}^0\right] = -\frac{\hbar^2}{2m_e}\left(i\vec{\nabla} + \frac{e}{\hbar c}\vec{A}(\vec{r})\right)^2 \hat{\Psi}_e^{0+}(\vec{r}) - \quad (21)$$

$$-\int d^2\vec{r}' V_{Coul.}(\vec{r}-\vec{r}')\hat{\Psi}_e^{0+}(\vec{r})\hat{\rho}^0(\vec{r}'),$$

$$i\hbar \frac{d\hat{\Psi}_h^{0+}(\vec{r})}{dt} = -\frac{\hbar^2}{2m_h}\left(i\vec{\nabla} - \frac{e}{\hbar c}\vec{A}(\vec{r})\right)^2 \hat{\Psi}_h^{0+}(\vec{r}) + \int d^2\vec{r}' V_{Coul.}(\vec{r}-\vec{r}')\hat{\Psi}_h^{0+}(\vec{r})\hat{\rho}^0(\vec{r}');$$

$$\hat{\rho}^{0+}(\vec{r}') = \hat{\rho}^0(\vec{r}').$$

The time derivatives $d\left(\hat{\rho}_i^0(\vec{r})\right)/dt$ and the continuity equations were derived using equations of motion

$$\frac{d}{dt}\hat{\rho}_i^0(\vec{r}) = \frac{d}{dt}\left(\hat{\Psi}_i^{0+}(\vec{r})\hat{\Psi}_i^0(\vec{r})\right) = -\vec{\nabla}\hat{\vec{J}}_i^0(\vec{r}); \quad i = e, h.$$

$$\hat{\vec{J}}_e^0(\vec{r}) = \frac{\hbar}{2m_e i}\left(\hat{\Psi}_e^{0+}(\vec{r})\vec{\nabla}\hat{\Psi}_e^0(\vec{r}) - \vec{\nabla}\hat{\Psi}_e^{0+}(\vec{r})\hat{\Psi}_e^0(\vec{r})\right) +$$

$$+\frac{e}{m_e c}\vec{A}(\vec{r})\hat{\rho}_e^0(\vec{r}),$$

$$\hat{\vec{J}}_h^0(\vec{r}) = \frac{\hbar}{2m_h i}\left(\hat{\Psi}_h^{0+}(\vec{r})\vec{\nabla}\hat{\Psi}_h^0(\vec{r}) - \vec{\nabla}\hat{\Psi}_h^{0+}(\vec{r})\hat{\Psi}_h^0(\vec{r})\right) - \quad (22)$$

$$-\frac{e}{m_h c}\vec{A}(\vec{r})\hat{\rho}_h^0(\vec{r}).$$

The equations of motion for the dressed field operators $\hat{\Psi}_i^+(\vec{r})$ and $\hat{\Psi}_i(\vec{r})$ can be obtained taking into account the equation (21) and time derivatives of the unitary transformation operators $\hat{u}(\vec{r}) = \hat{u}^0(\vec{r})$ and $\hat{u}^+(\vec{r}) = \hat{u}^{0+}(\vec{r})$, which coincide in the bare and in the dressed representations as well as operators $\hat{\rho}_i(\vec{r}) = \hat{\rho}_i^0(\vec{r})$, $\hat{\omega}_i(\vec{r}) = \hat{\omega}_i^0(\vec{r})$. Therefore, in calculations can be used one of them without supplementary specifications.

Following procedure used in Ref [9], we obtain two equations

$$\frac{i\hbar d\hat{\Psi}_e(\vec{r})}{dt} = i\hbar \frac{d}{dt}\left(\hat{u}^+(\vec{r})\hat{\Psi}_e^0(\vec{r})\right) = i\hbar \frac{d}{dt}\hat{u}^+(\vec{r})\cdot\hat{\Psi}_e^0(\vec{r}) + \hat{u}^+(\vec{r})\frac{i\hbar d}{dt}\hat{\Psi}_e^0(\vec{r}),$$

$$\frac{i\hbar d\hat{\Psi}_h(\vec{r})}{dt} = i\hbar \frac{d}{dt}\left(\hat{u}(\vec{r})\hat{\Psi}_h^0(\vec{r})\right) = i\hbar \frac{d}{dt}\hat{u}(\vec{r})\cdot\hat{\Psi}_h^0(\vec{r}) + \hat{u}(\vec{r})\frac{i\hbar d}{dt}\hat{\Psi}_h^0(\vec{r}). \tag{23}$$

To obtain the time derivatives of the unitary transformation operators $\exp\left(\pm\frac{ie}{\hbar c}\hat{\omega}(\vec{r})\right)$, we take into account that operators $d\hat{\omega}(\vec{r})/dt$ and $\hat{\omega}(\vec{r})$ do not commute. This was pointed out by Jackiw and Pi in Ref. [1], and demonstrated in Ref. [9] as follows

$$\left[\frac{d\hat{\omega}(\vec{r})}{dt},\hat{\omega}(\vec{r})\right] = -i\hat{L}(\vec{r}) = \left(\frac{\phi e}{\alpha}\right)^2 \int d^2\vec{r}' \int d^2\vec{r}'' \theta(\vec{r}-\vec{r}') \times$$
$$\times \theta(\vec{r}-\vec{r}'')\left[\frac{d\hat{\rho}(\vec{r}')}{dt},\hat{\rho}(\vec{r}'')\right] = \left(\frac{\phi e}{\alpha}\right)^2 \int d^2\vec{r}' \int d^2\vec{r}'' \theta(\vec{r}-\vec{r}')\theta(\vec{r}-\vec{r}'') \times$$
$$\times \left\{\left[\frac{d\hat{\rho}_e(\vec{r}')}{dt},\hat{\rho}_e(\vec{r}'')\right] + \left[\frac{d\hat{\rho}_h(\vec{r}')}{dt},\hat{\rho}_h(\vec{r}'')\right]\right\} = \left(\frac{\phi e}{\alpha}\right)^2 \int d^2\vec{r}' \int d^2\vec{r}'' \times \tag{24}$$
$$\times \vec{\nabla}'\theta(\vec{r}-\vec{r}')\theta(\vec{r}-\vec{r}'')\left\{\left[\hat{\vec{J}}_e(\vec{r}'),\hat{\rho}_e(\vec{r}'')\right] + \left[\hat{\vec{J}}_h(\vec{r}')\hat{\rho}_h(\vec{r}'')\right]\right\}$$

where we used expressions (22), what leads to the form

$$\hat{L}(\vec{r}) = \frac{\hbar}{2m_e}\hat{M}_e(\vec{r}) + \frac{\hbar}{2m_h}\hat{M}_h(\vec{r}),$$
$$\hat{M}_i(\vec{r}) = \left(\frac{\phi e}{\alpha}\right)^2 \int d^2\vec{r}' \int d^2\vec{r}'' \vec{\nabla}'\theta(\vec{r}-\vec{r}')\theta(\vec{r}-\vec{r}'') \times \tag{25}$$
$$\times \left\{\left[\hat{\Psi}_i^{0+}(\vec{r}')\vec{\nabla}'\hat{\Psi}_i^0(\vec{r}'),\hat{\rho}_i^0(\vec{r}'')\right] - \left[\vec{\nabla}'\hat{\Psi}_i^{0+}(\vec{r}')\hat{\Psi}_i^0(\vec{r}'),\hat{\rho}_i^0(\vec{r}'')\right]\right\}; i=e,h.$$

The commutation relations, which were substituted into formulas (25), were calculated in Ref [9] and look as follows

$$\left[\hat{\Psi}_i^{0+}(\vec{r}')\vec{\nabla}'\hat{\Psi}_i^0(\vec{r}'),\hat{\rho}_i^0(\vec{r}'')\right] = \hat{\Psi}_i^{0+}(\vec{r}')\vec{\nabla}'\left(\delta^2(\vec{r}'-\vec{r}'')\hat{\Psi}_i^0(\vec{r}'')\right) -$$
$$-\delta^2(\vec{r}'-\vec{r}'')\hat{\Psi}_i^{0+}(\vec{r}'')\vec{\nabla}'\hat{\Psi}_i^0(\vec{r}'),$$
$$\left[\vec{\nabla}'\hat{\Psi}_i^{0+}(\vec{r}')\hat{\Psi}_i^0(\vec{r}'),\hat{\rho}_i^0(\vec{r}'')\right] = -\vec{\nabla}'\left(\delta^2(\vec{r}'-\vec{r}'')\hat{\Psi}_i^{0+}(\vec{r}'')\right)\hat{\Psi}_i^0(\vec{r}') + \tag{26}$$
$$+\vec{\nabla}'\hat{\Psi}_i^{0+}(\vec{r}')\delta^2(\vec{r}'-\vec{r}'')\hat{\Psi}_i^0(\vec{r}''), i=e,h.$$

The properties $\Delta'\theta(\vec{r}-\vec{r}')=0$ and $(\vec{\nabla}'\theta(\vec{r}-\vec{r}'))^2=|\vec{r}-\vec{r}'|^{-2}$ help to effectuate the next calculations, which lead to expressions

$$\hat{M}_i(\vec{r})=2\left(\frac{\phi e}{\alpha}\right)^2\int d^2\vec{r}'\frac{\hat{\rho}_i^0(\vec{r}')}{|\vec{r}-\vec{r}'|^2}, i=e,h;$$

$$\hat{L}(\vec{r})=\left(\frac{\phi e}{\alpha}\right)\hbar\left[\frac{1}{m_e}\int d^2\vec{r}'\frac{\hat{\rho}_e^0(\vec{r}')}{|\vec{r}-\vec{r}'|^2}+\frac{1}{m_h}\int d^2\vec{r}'\frac{\hat{\rho}_h^0(\vec{r}')}{|\vec{r}-\vec{r}'|^2}\right],$$ (27)

$$[\hat{L}(\vec{r}),\hat{\omega}(\vec{r})]=[\hat{L}(\vec{r}),\hat{\rho}_i(\vec{r})]=0.$$

To derive time derivatives of the unitary transformation operators we will use the following formula

$$\frac{d}{dt}e^{\pm\frac{ie}{\hbar c}\hat{\omega}(\vec{r})}=\sum_{n=0}^{\infty}\left(\pm\frac{ie}{\hbar c}\right)^n\frac{1}{n!}\frac{d}{dt}\hat{\omega}^n(\vec{r}),$$ (28)

The first steps are the following results

$$\frac{d}{dt}\hat{\omega}^2(\vec{r})=2\hat{\omega}(\vec{r})\frac{d\hat{\omega}(\vec{r})}{dt}-i\hat{L}(\vec{r})=2\frac{d\hat{\omega}(\vec{r})}{dt}\hat{\omega}(\vec{r})+i\hat{L}(\vec{r}),$$

$$\frac{d}{dt}\hat{\omega}^3(\vec{r})=3\hat{\omega}^2(\vec{r})\frac{d\hat{\omega}(\vec{r})}{dt}-3i\hat{L}(\vec{r})\hat{\omega}(\vec{r})=3\frac{d\hat{\omega}(\vec{r})}{dt}\hat{\omega}^2(\vec{r})+3i\hat{L}(\vec{r})\hat{\omega}(\vec{r}),$$

$$\frac{d}{dt}\hat{\omega}^n(\vec{r})=n\hat{\omega}^{(n-1)}(\vec{r})\frac{d\hat{\omega}(\vec{r})}{dt}-iX_n\hat{L}(\vec{r})\hat{\omega}^{(n-2)}(\vec{r})=$$

$$=n\frac{d\hat{\omega}(\vec{r})}{dt}\hat{\omega}^{(n-1)}(\vec{r})+iX_n\hat{L}(\vec{r})\hat{\omega}^{(n-2)}(\vec{r}),$$ (29)

$$X_n=(n-1)+X_{n-1}+\frac{n(n-1)}{2}, n\geq 2.$$

The last equation was used, as follows

$$\frac{d}{dt}e^{\pm\frac{ie}{\hbar c}\hat{\omega}(\vec{r})}=\left(\pm\frac{ie}{\hbar c}\right)\frac{d\hat{\omega}(\vec{r})}{dt}+\sum_{n=2}^{\infty}\left(\pm\frac{ie}{\hbar c}\right)^n\frac{1}{n!}\frac{d}{dt}\hat{\omega}(\vec{r})^n=$$

$$=\left(\pm\frac{ie}{\hbar c}\right)\frac{d\hat{\omega}(\vec{r})}{dt}+\sum_{k=1}^{\infty}\left(\pm\frac{ie}{\hbar c}\right)^{(k+1)}\frac{1}{k!}\hat{\omega}^k(\vec{r})\frac{d\hat{\omega}(\vec{r})}{dt}-$$

$$-i\hat{L}(\vec{r})\sum_{m=0}^{\infty}\left(\pm\frac{ie}{\hbar c}\right)^{(m+2)}\frac{X_{(m+2)}}{(m+2)!}\hat{\omega}^m(\vec{r})=$$

$$
\begin{aligned}
&= \left(\pm\frac{ie}{\hbar c}\right)\frac{d\hat{\omega}(\vec{r})}{dt} + \frac{d\hat{\omega}(\vec{r})}{dt}\sum_{k=1}^{\infty}\left(\pm\frac{ie}{\hbar c}\right)^{(k+1)}\frac{1}{k!}\hat{\omega}^k(\vec{r}) + \\
&+ i\hat{L}(\vec{r})\sum_{m=0}^{\infty}\left(\pm\frac{ie}{\hbar c}\right)^{(m+2)}\frac{X_{(m+2)}}{(m+2)!}\hat{\omega}^m(\vec{r}) = \\
&= e^{\pm\frac{ie}{\hbar c}\hat{\omega}(\vec{r})}\left[\left(\pm\frac{ie}{\hbar c}\right)\frac{d\hat{\omega}(\vec{r})}{dt} + \frac{i}{2}\left(\frac{e}{\hbar c}\right)^2\hat{L}(\vec{r})\right] = \\
&= \left[\left(\pm\frac{ie}{\hbar c}\right)\frac{d\hat{\omega}(\vec{r})}{dt} - \frac{i}{2}\left(\frac{e}{\hbar c}\right)^2\hat{L}(\vec{r})\right]e^{\pm\frac{ie}{\hbar c}\hat{\omega}(\vec{r})}. \\
&\cdot \frac{X_{(m+2)}}{(m+2)!} = \frac{1}{2}\cdot\frac{1}{m!}, m \geq 0.
\end{aligned}
\quad (30)
$$

Now we can calculate the commutation relation between the time derivative $d\hat{\omega}(\vec{r})/dt$ and the unitary transformation operators $e^{\pm\frac{ie}{\hbar c}\hat{\omega}(\vec{r})}$:

$$\left[\frac{d\hat{\omega}(\vec{r})}{dt}, e^{\pm\frac{ie}{\hbar c}\hat{\omega}(\vec{r})}\right] = \sum_{n=0}^{\infty}\left(\pm\frac{ie}{\hbar c}\right)^n\cdot\frac{1}{n!}\left[\frac{d\hat{\omega}(\vec{r})}{dt}, \hat{\omega}^n(\vec{r})\right]. \quad (31)$$

To clarify this issue, we recall the basic equalities

$$\left[\frac{d\hat{\omega}(\vec{r})}{dt}, \hat{\omega}(\vec{r})\right] = -i\hat{L}(\vec{r}), \quad \left[\frac{d\hat{\omega}^2(\vec{r})}{dt}, \hat{\omega}(\vec{r})\right] = -2i\hat{L}(\vec{r})\hat{\omega}(\vec{r}).$$

They lead to the recurrent formula

$$\left[\frac{d\hat{\omega}(\vec{r})}{dt}, \hat{\omega}^n(\vec{r})\right] = -in\hat{L}(\vec{r})\hat{\omega}^{n-1}(\vec{r}), \quad (32)$$

which solves the problem:

$$\left[\frac{d\hat{\omega}(\vec{r})}{dt}, e^{\pm\frac{ie}{\hbar c}\hat{\omega}(\vec{r})}\right] = \left(\pm\frac{ie}{\hbar c}\right)\hat{L}(\vec{r})e^{\pm\frac{ie}{\hbar c}\hat{\omega}(\vec{r})}. \quad (33)$$

Now, to derive the equations of motion for the dressed field operators $\hat{\Psi}_i(\vec{r})$ and $\hat{\Psi}_i^+(\vec{r})$ we consider equations

$$i\hbar \frac{d}{dt}\hat{\Psi}_e(\vec{r}) = i\hbar \frac{d}{dt}\left(\hat{u}^+(\vec{r})\hat{\Psi}_e^0(\vec{r})\right) = i\hbar \frac{d\hat{u}^+(\vec{r})}{dt}\hat{\Psi}_e^0(\vec{r}) + \hat{u}^+(\vec{r})i\hbar \frac{d\hat{\Psi}_e^0(\vec{r})}{dt},$$
$$i\hbar \frac{d}{dt}\hat{\Psi}_h(\vec{r}) = i\hbar \frac{d}{dt}\left(\hat{u}(\vec{r})\hat{\Psi}_h^0(\vec{r})\right) = i\hbar \frac{d\hat{u}(\vec{r})}{dt}\hat{\Psi}_h^0(\vec{r}) + \hat{u}(\vec{r})i\hbar \frac{d\hat{\Psi}_h^0(\vec{r})}{dt}.$$
(34)

We recall that the operators $\hat{\omega}(\vec{r})$ and $d\hat{\omega}(\vec{r})/dt$ do not commute and their commutation equals to

$$\left[\frac{d\hat{\omega}(\vec{r})}{dt}, \hat{\omega}(\vec{r})\right] = -i\hat{L}(\vec{r}),$$
(35)

$$\hat{L}^+(\vec{r}) = \hat{L}(\vec{r}),\ \left[\hat{L}(\vec{r}),\hat{\rho}_i(\vec{r})\right] = 0,\ \left[\hat{L}(\vec{r}),\hat{\omega}(\vec{r}')\right] = 0.$$

Similar to as it was done in derivation of Eq. (30) we can write

$$i\hbar \frac{d\hat{u}(\vec{r})}{dt} = \hat{u}(\vec{r})\left[-\frac{e}{c}\frac{d\hat{\omega}(\vec{r})}{dt} - \frac{e^2}{2\hbar c^2}\hat{L}(\vec{r})\right] =$$
$$= \left[-\frac{e}{c}\frac{d\hat{\omega}(\vec{r})}{dt} + \frac{e^2}{2\hbar c^2}\hat{L}(\vec{r})\right]\hat{u}(\vec{r}),$$
$$\hat{u}(\vec{r}) = e^{\frac{ie}{\hbar c}\hat{\omega}(\vec{r})},$$
$$i\hbar \frac{d\hat{u}^+(\vec{r})}{dt} = \hat{u}^+(\vec{r})\left[\frac{e}{c}\frac{d\hat{\omega}(\vec{r})}{dt} - \frac{e^2}{2\hbar c^2}\hat{L}(\vec{r})\right] =$$
$$= \left[\frac{e}{c}\frac{d\hat{\omega}(\vec{r})}{dt} + \frac{e^2}{2\hbar c^2}\hat{L}(\vec{r})\right]\hat{u}^+(\vec{r}),$$
(36)
$$\hat{u}^+(\vec{r}) = e^{-\frac{ie}{\hbar c}\hat{\omega}(\vec{r})}.$$

Below we will use formulas describing the action of the operator $\left(-i\vec{\nabla}\right)$ on the unitary transformation operators $e^{\pm\frac{ie}{\hbar c}\hat{\omega}(\vec{r})}$:

$$\left(-i\vec{\nabla}\right)e^{\pm\frac{ie}{\hbar c}\hat{\omega}(\vec{r})} = \pm\frac{e}{\hbar c}\hat{\vec{a}}(\vec{r})e^{\pm\frac{ie}{\hbar c}\hat{\omega}(\vec{r})};\quad \hat{\vec{a}}(\vec{r}) = \vec{\nabla}\hat{\omega}(\vec{r}),$$

$$e^{-\frac{ie}{\hbar c}\hat{\omega}(\vec{r})}\left(-i\vec{\nabla}+\frac{e}{\hbar c}\vec{A}(\vec{r})\right)^2 \hat{\Psi}_e^0(\vec{r}) = \left(-i\vec{\nabla}+\frac{e}{\hbar c}\vec{A}(\vec{r})+\frac{e}{\hbar c}\hat{\vec{a}}(\vec{r})\right)^2 \hat{\Psi}_e(\vec{r}),$$

$$e^{\frac{ie}{\hbar c}\hat{\omega}(\vec{r})}\left(-i\vec{\nabla}-\frac{e}{\hbar c}\vec{A}(\vec{r})\right)^2 \hat{\Psi}_h^0(\vec{r}) = \left(-i\vec{\nabla}-\frac{e}{\hbar c}\vec{A}(\vec{r})-\frac{e}{\hbar c}\hat{\vec{a}}(\vec{r})\right)^2 \hat{\Psi}_h(\vec{r}). \quad (37)$$

$$\hat{\Psi}_e(\vec{r})=\hat{u}^+(\vec{r})\hat{\Psi}_e^0(\vec{r}),\hat{\Psi}_h(\vec{r})=\hat{u}(\vec{r})\hat{\Psi}_e^0(\vec{r}),\left[\vec{a}(\vec{r}),\hat{\omega}(\vec{r}')\right]=0.$$

The time derivatives of the dressed field operators can be expressed through the derivatives of their components:

$$i\hbar\frac{d}{dt}\hat{\Psi}_e(\vec{r}) = i\hbar\frac{d\hat{u}^+(\vec{r})}{dt}\hat{\Psi}_e^0(\vec{r}) + \hat{u}^+(\vec{r})i\hbar\frac{d\hat{\Psi}_e^0(\vec{r})}{dt},$$

$$i\hbar\frac{d}{dt}\hat{\Psi}^h(\vec{r}) = i\hbar\frac{d\hat{u}(\vec{r})}{dt}\hat{\Psi}_h^0(\vec{r}) + \hat{u}(\vec{r})i\hbar\frac{d\hat{\Psi}_h^0(\vec{r})}{dt}. \quad (38)$$

Using formulas (21), (36) and (37) we obtain

$$i\hbar\frac{d\hat{\Psi}_e(\vec{r})}{dt} = \frac{\hbar^2}{2m_e}\left(-i\vec{\nabla}-\frac{e}{\hbar c}\vec{A}(\vec{r})+\frac{e}{\hbar c}\hat{\vec{a}}(\vec{r})\right)^2 \hat{\Psi}_e(\vec{r})+$$

$$+\frac{e}{c}\frac{d\hat{\omega}(\vec{r})}{dt}\hat{\Psi}_e(\vec{r})+\frac{e^2}{2\hbar c^2}\hat{L}(\vec{r})\hat{\Psi}_e(\vec{r})+\int d^2\vec{r}'V_{Coul.}(\vec{r}-\vec{r}')\hat{\rho}(\vec{r}')\hat{\Psi}_e(\vec{r}),$$

$$i\hbar\frac{d\hat{\Psi}_h(\vec{r})}{dt} = \frac{\hbar^2}{2m_h}\left(-i\vec{\nabla}-\frac{e}{\hbar c}\vec{A}(\vec{r})-\frac{e}{\hbar c}\hat{\vec{a}}(\vec{r})\right)^2 \hat{\Psi}_h(\vec{r})-$$

$$-\frac{e}{c}\frac{d\hat{\omega}(\vec{r})}{dt}\hat{\Psi}_h(\vec{r})+\frac{e^2}{2\hbar c^2}\hat{L}(\vec{r})\hat{\Psi}_h(\vec{r})-\int d^2\vec{r}'V_{Coul.}(\vec{r}-\vec{r}')\hat{\rho}(\vec{r}')\hat{\Psi}_h(\vec{r}), \quad (39)$$

$$\hat{\rho}(\vec{r}')=\hat{\rho}_e(\vec{r}')-\hat{\rho}_h(\vec{r}');\hat{\rho}_i(\vec{r}')=\hat{\Psi}_i^+(\vec{r}')\hat{\Psi}_i(\vec{r}'),i=e,h.$$

Unlike equations of motion (21) for the bare field operators $\hat{\Psi}_i^0(\vec{r})$, equations (39) contain new operators such as $\hat{\omega}(\vec{r})$, $\dfrac{\hat{\omega}(\vec{r})}{dt}$ and $\vec{\nabla}\hat{\omega}(\vec{r})=\hat{\vec{a}}(\vec{r})$, which characterize the Chern-Simons gauge field. Similar to the case of the one-component two-dimensional electron gas (2DEG), in the two-component 2D e-h system the quantum point vortices also give rise to the vector-potential $\vec{a}(\vec{r})$ and to the scalar potential $\dfrac{1}{c}\dfrac{d\hat{\omega}}{dt}$, which appear supplementary to the vector-potential $\vec{A}(\vec{r})$ of the external magnetic field. But, as it was mentioned above, they depend on the difference of the density operators $\hat{\rho}(\vec{r}) = \hat{\rho}_e(\vec{r}) - \hat{\rho}_h(\vec{r})$ due to different signs of the electrical charges of electrons and holes.

The vector potential $\vec{a}(\vec{r})$ in our case cannot compensate the vector potential $\vec{A}(\vec{r})$ created by the external magnetic field. The C-S vector potential $\vec{a}(\vec{r})$ vanishes in the mean-field approximation if the average densities $\langle \rho_e \rangle$ and $\langle \rho_h \rangle$ coincide. Nevertheless, the numbers $\phi$ of the quantum point vortices attached to each electron and to each hole can be different from zero and the composite particles in the zero-order approximation will be subjected only to the external magnetic field. In the zero-order approximation they undergo the Landau quantization under the influence of the external magnetic field and will undergo perturbations in the next orders of the perturbation theory. Side by side with the equations of motion (39) giving the time evolution of the dressed field operators we need the Hamiltonian describing the 2D e-h system in the presence of the Chern-Simons gauge field. It can be easily obtained substituting in the Hamiltonian (19) the bare field operators $\hat{\Psi}_i^{0+}(\vec{r})$ and $\hat{\Psi}_i^0(\vec{r})$ by the dressed field operators

$$\hat{\Psi}_e^0(\vec{r}) = u(\vec{r})\hat{\Psi}_e(\vec{r}),\ \hat{\Psi}_h^0(\vec{r}) = u^+(\vec{r})\hat{\Psi}_h(\vec{r}),\ \hat{\Psi}_e^{0+}(\vec{r}) = \hat{\Psi}_e^+(\vec{r})\hat{u}^+(\vec{r}),$$

$$\hat{\Psi}_h^{0+}(\vec{r}) = \hat{\Psi}_h^+(\vec{r})\hat{u}(\vec{r}),\ \hat{u}(\vec{r}) = e^{\frac{ie}{\hbar c}\hat{\omega}(\vec{r})},\ \hat{u}^+(\vec{r}) = e^{-\frac{ie}{\hbar c}\hat{\omega}(\vec{r})}. \quad (40)$$

For example, the first two terms of the Hamiltonian (19) is transformed as follows

$$\hat{K}^0 = \frac{\hbar^2}{2m_e}\int d^2\vec{r}'\hat{\Psi}_e^{0+}(\vec{r}')\left(-i\vec{\nabla}' + \frac{e}{\hbar c}\vec{A}(\vec{r})\right)^2 \hat{\Psi}_e^0(\vec{r}') +$$

$$+ \frac{\hbar^2}{2m_h}\int d^2\vec{r}'\hat{\Psi}_h^{0+}(\vec{r}')\left(-i\vec{\nabla}' - \frac{e}{\hbar c}\vec{A}(\vec{r})\right)^2 \hat{\Psi}_h^0(\vec{r}') =$$

$$= \frac{\hbar^2}{2m_e}\int d^2\vec{r}'\hat{\Psi}_e^+(\vec{r}')e^{-\frac{ie}{\hbar c}\hat{\omega}(\vec{r}')}\left(-i\vec{\nabla}' + \frac{e}{\hbar c}\vec{A}(\vec{r}')\right)^2 e^{\frac{ie}{\hbar c}\hat{\omega}(\vec{r}')}\hat{\Psi}_e(\vec{r}') +$$

$$+ \frac{\hbar^2}{2m_h}\int d^2\vec{r}'\hat{\Psi}_h^+(\vec{r}')e^{\frac{ie}{\hbar c}\hat{\omega}(\vec{r}')}\left(-i\vec{\nabla}' - \frac{e}{\hbar c}\vec{A}(\vec{r}')\right)^2 e^{-\frac{ie}{\hbar c}\hat{\omega}(\vec{r}')}\hat{\Psi}_h^0(\vec{r}') =$$

$$= \frac{\hbar^2}{2m_e}\int d^2\vec{r}'\hat{\Psi}_e^+(\vec{r}')\left(-i\vec{\nabla}' + \frac{e}{\hbar c}\vec{A}(\vec{r}') + \frac{e}{\hbar c}\hat{\vec{a}}(\vec{r}')\right)^2 \hat{\Psi}_e(\vec{r}') + \quad , \quad (41)$$

$$+ \frac{\hbar^2}{2m_h}\int d^2\vec{r}'\hat{\Psi}_h^+(\vec{r}')\left(-i\vec{\nabla}' - \frac{e}{\hbar c}\vec{A}(\vec{r}') - \frac{e}{\hbar c}\hat{\vec{a}}(\vec{r}')\right)^2 \hat{\Psi}_h(\vec{r}') = \hat{K}$$

$$\hat{\vec{a}}(\vec{r}') = \vec{\nabla}'\hat{\omega}(\vec{r}').$$

where we have used relations (37)

$$e^{-\frac{ie}{\hbar c}\hat{\omega}(\vec{r}')}\left(-i\vec{\nabla}' + \frac{e}{\hbar c}\vec{A}(\vec{r}')\right)^2 = \left(-i\vec{\nabla}' + \frac{e}{\hbar c}\vec{A}(\vec{r}') + \frac{e}{\hbar c}\hat{\vec{a}}(\vec{r}')\right)^2 e^{-\frac{ie}{\hbar c}\hat{\omega}(\vec{r}')},$$

$$e^{\frac{ie}{\hbar c}\hat{\omega}(\vec{r}')}\left(-i\vec{\nabla}' - \frac{e}{\hbar c}\vec{A}(\vec{r}')\right)^2 = \left(-i\vec{\nabla}' - \frac{e}{\hbar c}\vec{A}(\vec{r}') - \frac{e}{\hbar c}\hat{\vec{a}}(\vec{r}')\right)^2 e^{\frac{ie}{\hbar c}\hat{\omega}(\vec{r}')}.$$

(42)

The Hamiltonian $\hat{H}^0_{Coul.}$ of the Coulomb interaction of the bare electrons and holes can be transcribed in the dressed field operator representation, taking into account that

$$\hat{\rho}^0_i(\vec{r}) = \hat{\Psi}^{0+}_i(\vec{r})\hat{\Psi}^0_i(\vec{r}) = \hat{\rho}(\vec{r}) = \hat{\Psi}^+_i(\vec{r})\hat{\Psi}_i(\vec{r}), i = e, h.,$$
$$\hat{H}^0_{Coul.} = \hat{H}_{Coul.}.$$

(43)

The Hamiltonian of the e-h system in the presence of the C-S gauge field looks as

$$\hat{H} = \hat{K} + \hat{H}_{Coul.} = \frac{\hbar^2}{2m_e}\int d^2\vec{r}'\hat{\Psi}^+_e(\vec{r}')\left(-i\vec{\nabla}' + \frac{e}{\hbar c}\vec{A}(\vec{r}') + \frac{e}{\hbar c}\hat{\vec{a}}(\vec{r}')\right)^2 \hat{\Psi}_e(\vec{r}') +$$
$$+ \frac{\hbar^2}{2m_h}\int d^2\vec{r}'\hat{\Psi}^+_h(\vec{r}')\left(-i\vec{\nabla}' - \frac{e}{\hbar c}\vec{A}(\vec{r}') - \frac{e}{\hbar c}\hat{\vec{a}}(\vec{r}')\right)^2 \hat{\Psi}_h(\vec{r}') +$$
$$+ \frac{1}{2}\int d^2\vec{r}'\int d^2\vec{r}''V_{Coul.}(\vec{r}' - \vec{r}'')\hat{\Psi}^+_e(\vec{r}')\hat{\rho}_e(\vec{r}'')\hat{\Psi}_e(\vec{r}') +$$
$$+ \frac{1}{2}\int d^2\vec{r}'\int d^2\vec{r}''V_{Coul.}(\vec{r}' - \vec{r}'')\hat{\Psi}^+_h(\vec{r}')\hat{\rho}_h(\vec{r}'')\hat{\Psi}_h(\vec{r}') -$$
$$- \int d^2\vec{r}'\int d^2\vec{r}''V_{Coul.}(\vec{r}' - \vec{r}'')\hat{\Psi}^+_e(\vec{r}')\hat{\rho}_h(\vec{r}'')\hat{\Psi}_e(\vec{r}').$$

(44)

The Hamiltonian (44) is much more complicated than its bare counterpart (19), because it contains a nonlinear form the vector potential operator $\hat{\vec{a}}(\vec{r}')$ created by the C-S gauge field. To obtain the equations of motion for the dressed field operators we start with the Schrodinger equations for electrons and holes

$$i\hbar\frac{d\hat{\Psi}_i(\vec{r})}{dt} = \left[\hat{\Psi}_i(\vec{r}), \hat{H}\right], \quad i = e, h$$

(45)

In this case we need to calculate commutation relations of the field operators with the operator $\hat{\vec{a}}(\vec{r}')$

$$\left[\hat{\Psi}_e(\vec{r}),\hat{\vec{a}}(\vec{r}')\frac{e}{\hbar c}\right]=-\phi\vec{\nabla}'\theta(\vec{r}'-\vec{r})\hat{\Psi}_e(\vec{r}),$$

$$\left[\hat{\Psi}_h(\vec{r}),\hat{\vec{a}}(\vec{r}')\frac{e}{\hbar c}\right]=\phi\vec{\nabla}'\theta(\vec{r}'-\vec{r})\hat{\Psi}_h(\vec{r}),$$

$$\left[\hat{\Psi}_e^+(\vec{r}),\hat{\vec{a}}(\vec{r}')\frac{e}{\hbar c}\right]=\phi\vec{\nabla}'\theta(\vec{r}'-\vec{r})\hat{\Psi}_e^+(\vec{r}),$$

$$\left[\hat{\Psi}_h^+(\vec{r}),\hat{\vec{a}}(\vec{r}')\frac{e}{\hbar c}\right]=-\phi\vec{\nabla}'\theta(\vec{r}'-\vec{r})\hat{\Psi}_h^+(\vec{r}),$$

$$\left[\hat{\Psi}_i(\vec{r}),\hat{\vec{a}}(\vec{r})\right]=\left[\hat{\Psi}_i^+(\vec{r}),\hat{\vec{a}}(\vec{r})\right]=0,\ i=e-h,$$

$$\left.\vec{\nabla}'\theta(\vec{r}'-\vec{r})\right|_{\vec{r}-\vec{r}'}=0.$$

(46)

Using these equations, we can write

$$\hat{\Psi}_e(\vec{r})\left(-i\vec{\nabla}'+\frac{e}{\hbar c}\vec{A}(\vec{r}')+\frac{e}{\hbar c}\hat{\vec{a}}(\vec{r}')\right)^2=$$

$$=\left(-i\vec{\nabla}'+\frac{e}{\hbar c}\vec{A}(\vec{r}')+\frac{e}{\hbar c}\hat{\vec{a}}(\vec{r}')-\phi\vec{\nabla}'\theta(\vec{r}'-\vec{r})\right)^2\hat{\Psi}_e(\vec{r}),$$

$$\hat{\Psi}_h(\vec{r})\left(-i\vec{\nabla}'-\frac{e}{\hbar c}\vec{A}(\vec{r}')-\frac{e}{\hbar c}\hat{\vec{a}}(\vec{r}')\right)^2=$$

$$=\left(-i\vec{\nabla}'-\frac{e}{\hbar c}\vec{A}(\vec{r}')-\frac{e}{\hbar c}\hat{\vec{a}}(\vec{r}')-\phi\vec{\nabla}'\theta(\vec{r}'-\vec{r})\right)^2\hat{\Psi}_h(\vec{r}),$$

$$\hat{\Psi}_e^+(\vec{r})\left(-i\vec{\nabla}'+\frac{e}{\hbar c}\vec{A}(\vec{r}')+\frac{e}{\hbar c}\hat{\vec{a}}(\vec{r}')\right)^2=$$

$$=\left(-i\vec{\nabla}'+\frac{e}{\hbar c}\vec{A}(\vec{r}')+\frac{e}{\hbar c}\hat{\vec{a}}(\vec{r}')+\phi\vec{\nabla}'\theta(\vec{r}'-\vec{r})\right)^2\hat{\Psi}_e^+(\vec{r}),$$

(47)

$$\hat{\Psi}_h^+(\vec{r})\left(-i\vec{\nabla}'-\frac{e}{\hbar c}\vec{A}(\vec{r}')-\frac{e}{\hbar c}\hat{\vec{a}}(\vec{r}')\right)^2=$$

$$=\left(-i\vec{\nabla}'-\frac{e}{\hbar c}\vec{A}(\vec{r}')-\frac{e}{\hbar c}\hat{\vec{a}}(\vec{r}')+\phi\vec{\nabla}'\theta(\vec{r}'-\vec{r})\right)^2\hat{\Psi}_h^+(\vec{r}).$$

It should be remembered, that dressed field operators $\hat{\Psi}_i(\vec{r})$ and $\hat{\Psi}_i^+(\vec{r})$ obey the Fermi or Bose statistics with commutation relations

$$\hat{\Psi}_i(\vec{r})\hat{\Psi}_j^+(\vec{r})\pm\hat{\Psi}_j^+(\vec{r})\hat{\Psi}_i(\vec{r})=\delta_{ij}\delta^2(\vec{r}-\vec{r}'),\quad(F)$$
$$\hat{\Psi}_i(\vec{r})\hat{\Psi}_j^+(\vec{r})\pm\hat{\Psi}_j^+(\vec{r})\hat{\Psi}_i(\vec{r})=0,\quad\quad\quad(B)$$

(48)

and all the formulas obtained above are valid in both cases. For example, to derive the next commutation relations we have to use both formulas (47) and (48) as follows

$$\left[\hat{\Psi}_e(\vec{r}), \hat{\Psi}_e^+(\vec{r}')\left(-i\vec{\nabla}' + \frac{e}{\hbar c}\vec{A}(\vec{r}') + \frac{e}{\hbar c}\hat{\vec{a}}(\vec{r}')\right)^2 \hat{\Psi}_e(\vec{r}')\right] =$$

$$= \delta^2(\vec{r} - \vec{r}')\left(-i\vec{\nabla}' + \frac{e}{\hbar c}\vec{A}(\vec{r}') + \frac{e}{\hbar c}\hat{\vec{a}}(\vec{r}')\right)^2 \hat{\Psi}_e(\vec{r}') + \left(\phi\vec{\nabla}'\theta(\vec{r}' - \vec{r})\right)^2 \times$$

$$\times \hat{\rho}_e(\vec{r}')\hat{\Psi}_e(\vec{r}),$$

$$\left[\hat{\Psi}_e(\vec{r}), \frac{\hbar^2}{2m_e}\int d^2\vec{r}' \hat{\Psi}_e^+(\vec{r}')\left(-i\vec{\nabla}' + \frac{e}{\hbar c}\vec{A}(\vec{r}') + \frac{e}{\hbar c}\hat{\vec{a}}(\vec{r}')\right)^2 \hat{\Psi}_e(\vec{r}')\right] =$$

$$= \frac{\hbar^2}{2m_e}\left(-i\vec{\nabla} + \frac{e}{\hbar c}\vec{A}(\vec{r}) + \frac{e}{\hbar c}\hat{\vec{a}}(\vec{r})\right)^2 \hat{\Psi}_e(\vec{r}) +$$

$$+ \phi^2 \frac{\hbar^2}{2m_e}\int d^2\vec{r}' \frac{\hat{\rho}_e(\vec{r}')}{|\vec{r} - \vec{r}'|^2}\hat{\Psi}_e(\vec{r}) -$$

$$- \phi\frac{\hbar^2}{m_e}\int d^2\vec{r}' \vec{\nabla}'\theta(\vec{r}' - \vec{r})\hat{\Psi}_e^+(\vec{r}')\left(-i\vec{\nabla}' + \frac{e}{\hbar c}\vec{A}(\vec{r}') + \frac{e}{\hbar c}\hat{\vec{a}}(\vec{r}')\right)\hat{\Psi}_e(\vec{r}')\hat{\Psi}_e(r),$$

$$\left[\hat{\Psi}_h(\vec{r}), \frac{\hbar^2}{2m_h}\int d^2\vec{r}' \hat{\Psi}_h^+(\vec{r}')\left(-i\vec{\nabla}' - \frac{e}{\hbar c}\vec{A}(\vec{r}') - \frac{e}{\hbar c}\hat{\vec{a}}(\vec{r}')\right)^2 \hat{\Psi}_h(\vec{r}')\right] =$$

$$= \frac{\hbar^2}{2m_h}\left(-i\vec{\nabla} - \frac{e}{\hbar c}\vec{A}(\vec{r}) - \frac{e}{\hbar c}\hat{\vec{a}}(\vec{r})\right)^2 \hat{\Psi}_h(\vec{r}) +$$

$$+ \phi^2 \frac{\hbar^2}{2m_h}\int d^2\vec{r}' \frac{\hat{\rho}_h(\vec{r}')}{|\vec{r} - \vec{r}'|^2}\hat{\Psi}_h(\vec{r}) - \qquad (49)$$

$$- \phi\frac{\hbar^2}{m_h}\int d^2\vec{r}' \vec{\nabla}'\theta(\vec{r}' - \vec{r})\hat{\Psi}_h^+(\vec{r}')\left(-i\vec{\nabla}' - \frac{e}{\hbar c}\vec{A}(\vec{r}') - \frac{e}{\hbar c}\hat{\vec{a}}(\vec{r}')\right)\hat{\Psi}_h(\vec{r}')\hat{\Psi}_h(r).$$

Here we took into account the properties described by formulas (46) and the equalities written below

$$\vec{\nabla}'\vec{A}(\vec{r}') = \vec{\nabla}'\hat{\vec{a}}(\vec{r}') = 0; \quad \left[\hat{\Psi}_i(\vec{r}), \hat{\vec{a}}(\vec{r})\right] = 0;$$

$$\left(\vec{\nabla}'\theta(\vec{r}' - \vec{r})\right)^2 = \frac{1}{|\vec{r} - \vec{r}'|^2}; \quad \Delta'\theta(\vec{r}' - \vec{r}) = 0. \qquad (50)$$

Both integrals proportional to $\phi$ in Eq. (49) may be transformed introducing the dressed current density operators for the electrons $\hat{\vec{J}}_e(\vec{r})$ and for the holes $\hat{\vec{J}}_h(\vec{r})$ as well as the corresponding continuity equations as follows

$$\hat{\vec{J}}_e(\vec{r}) = \frac{\hbar^2}{2m_e i}\left(\hat{\Psi}_e^+(\vec{r})\vec{\nabla}\hat{\Psi}_e(\vec{r}) - \vec{\nabla}\hat{\Psi}_e^+(\vec{r})\hat{\Psi}_e(\vec{r})\right) +$$

$$+\left(\frac{e}{m_e c}\vec{A}(\vec{r}) + \frac{e}{m_e c}\hat{\vec{a}}(\vec{r})\right)\hat{\rho}_e(\vec{r}),$$

$$\hat{\vec{J}}_h(\vec{r}) = \frac{\hbar^2}{2m_h i}\left(\hat{\Psi}_h^+(\vec{r})\vec{\nabla}\hat{\Psi}_h(\vec{r}) - \vec{\nabla}\hat{\Psi}_h^+(\vec{r})\hat{\Psi}_h(\vec{r})\right) - \qquad (51)$$

$$-\left(\frac{e}{m_h c}\vec{A}(\vec{r}) + \frac{e}{m_h c}\hat{\vec{a}}(\vec{r})\right)\hat{\rho}_h(\vec{r}).$$

$$\frac{d\hat{\rho}_e(\vec{r})}{dt} = -\vec{\nabla}\hat{\vec{J}}_e(\vec{r}), \frac{d\hat{\rho}_h(\vec{r})}{dt} = -\vec{\nabla}\hat{\vec{J}}_h(\vec{r}).$$

The condition $\Delta'\theta(\vec{r}' - \vec{r}) = 0$ is useful for a simple integral transformation

$$-i\int d^2\vec{r}'\vec{\nabla}'\theta(\vec{r}' - \vec{r})\hat{\Psi}_i^+(\vec{r}')\vec{\nabla}'\hat{\Psi}_i(\vec{r}') =$$

$$= \frac{i}{2}\int d^2\vec{r}'\vec{\nabla}'\theta(\vec{r}' - \vec{r})\left(\nabla'\hat{\Psi}_i^+(\vec{r}')\hat{\Psi}_i(\vec{r}') - \hat{\Psi}_i^+(\vec{r}')\vec{\nabla}'\hat{\Psi}_i(\vec{r}')\right).$$

The obtained relations allow us to write the above-mentioned integrals in Eqs. (49) as follows

$$\phi\frac{\hbar^2}{m_e}\int d^2\vec{r}'\vec{\nabla}'\theta(\vec{r}' - \vec{r})\hat{\Psi}_e^+(\vec{r}')\left(-i\vec{\nabla}' + \frac{e}{\hbar c}\vec{A}(\vec{r}') + \frac{e}{\hbar c}\hat{\vec{a}}(\vec{r}')\right)\hat{\Psi}_e(\vec{r}')\hat{\Psi}_e(\vec{r}) =$$

$$= \phi\hbar\int d^2\vec{r}'\vec{\nabla}'\theta(\vec{r}' - \vec{r})\hat{\vec{J}}_e(\vec{r}')\hat{\Psi}_e(\vec{r}) = -\phi\hbar\int d^2\vec{r}'\theta(\vec{r}' - \vec{r})\times$$

$$\times\vec{\nabla}'\hat{\vec{J}}_e(\vec{r}')\hat{\Psi}_e(\vec{r}) = \phi\hbar\int d^2\vec{r}'\theta(\vec{r}' - \vec{r})\frac{d\hat{\rho}_e(\vec{r}')}{dt}\hat{\Psi}_e(\vec{r}) = -\frac{e}{c}\frac{d\hat{\omega}_e(\vec{r})}{dt}\hat{\Psi}_e(\vec{r}),$$

$$\phi\frac{\hbar^2}{m_h}\int d^2\vec{r}'\vec{\nabla}'\theta(\vec{r}' - \vec{r})\hat{\Psi}_h^+(\vec{r}')\left(-i\vec{\nabla}' - \frac{e}{\hbar c}\vec{A}(\vec{r}') - \frac{e}{\hbar c}\hat{\vec{a}}(\vec{r}')\right)\hat{\Psi}_h(\vec{r}')\hat{\Psi}_h(\vec{r}) = \qquad (52)$$

$$= \phi\hbar\int d^2\vec{r}'\vec{\nabla}'\theta(\vec{r}' - \vec{r})\hat{\vec{J}}_h(\vec{r}')\hat{\Psi}_h(\vec{r}) = -\phi\hbar\int d^2\vec{r}'\theta(\vec{r}' - \vec{r})\times$$

$$\times\vec{\nabla}'\hat{\vec{J}}_h(\vec{r}')\hat{\Psi}_h(\vec{r}) = \phi\hbar\int d^2\vec{r}'\theta(\vec{r}' - \vec{r})\frac{d\hat{\rho}_h(\vec{r}')}{dt}\hat{\Psi}_h(\vec{r}) = -\frac{e}{c}\frac{d\hat{\omega}_h(\vec{r})}{dt}\hat{\Psi}_h(\vec{r}),$$

$$\hat{\omega}_i(\vec{r}) = -\frac{\phi e}{\alpha}\int d^2\vec{r}'\theta(\vec{r} - \vec{r}')\hat{\rho}_i(\vec{r}'), i = e, h.$$

Combining the results expressed by formulas (49) - (52), we can formulate the main result of our calculations

$$\left[\hat{\Psi}_e(\vec{r}), \frac{\hbar^2}{2m_e}\int d^2\vec{r}'\hat{\Psi}_e^+(\vec{r}')\left(-i\vec{\nabla}' + \frac{e}{\hbar c}\vec{A}(\vec{r}') + \frac{e}{\hbar c}\hat{\vec{a}}(\vec{r}')\right)^2 \hat{\Psi}_e(\vec{r}')\right] =$$

$$= \frac{\hbar^2}{2m_e}\left(-i\vec{\nabla}' + \frac{e}{\hbar c}\vec{A}(\vec{r}) + \frac{e}{\hbar c}\hat{\vec{a}}(\vec{r})\right)^2 \hat{\Psi}_e(\vec{r}) + \phi^2 \frac{\hbar^2}{2m_e}\int d^2\vec{r}' \frac{\hat{\rho}_e(\vec{r}')}{|\vec{r}-\vec{r}'|^2} \times$$

$$\times \hat{\Psi}_e(\vec{r}) + \frac{e}{c}\frac{d\hat{\omega}_e(\vec{r})}{dt}\hat{\Psi}_e(\vec{r}),$$

$$\left[\hat{\Psi}_h(\vec{r}), \frac{\hbar^2}{2m_h}\int d^2\vec{r}'\hat{\Psi}_h^+(\vec{r}')\left(-i\vec{\nabla}' - \frac{e}{\hbar c}\vec{A}(\vec{r}') - \frac{e}{\hbar c}\hat{\vec{a}}(\vec{r}')\right)^2 \hat{\Psi}_h(\vec{r}')\right] =$$

$$= \frac{\hbar^2}{2m_h}\left(-i\vec{\nabla}' - \frac{e}{\hbar c}\vec{A}(\vec{r}) - \frac{e}{\hbar c}\hat{\vec{a}}(\vec{r})\right)^2 \hat{\Psi}_h(\vec{r}) + \phi^2 \frac{\hbar^2}{2m_h}\int d^2\vec{r}' \frac{\hat{\rho}_h(\vec{r}')}{|\vec{r}-\vec{r}'|^2} \times \qquad (53)$$

$$\times \hat{\Psi}_h(\vec{r}) + \frac{e}{c}\frac{d\hat{\omega}_h(\vec{r})}{dt}\hat{\Psi}_h(\vec{r}).$$

In a similar way one can derive the following commutations

$$\left[\hat{\Psi}_e(\vec{r}), \frac{\hbar^2}{2m_h}\int d^2\vec{r}'\hat{\Psi}_h^+(\vec{r}')\left(-i\vec{\nabla}' - \frac{e}{\hbar c}\vec{A}(\vec{r}') - \frac{e}{\hbar c}\hat{\vec{a}}(\vec{r}')\right)^2 \hat{\Psi}_h(\vec{r}')\right] =$$

$$= \phi^2 \frac{\hbar^2}{2m_h}\int d^2\vec{r}' \frac{\hat{\rho}_h(\vec{r}')}{|\vec{r}-\vec{r}'|^2}\hat{\Psi}_e(\vec{r}) - \frac{e}{c}\frac{d\hat{\omega}_h(\vec{r})}{dt}\hat{\Psi}_e(\vec{r}),$$

$$\left[\hat{\Psi}_h(\vec{r}), \frac{\hbar^2}{2m_e}\int d^2\vec{r}'\hat{\Psi}_h^+(\vec{r}')\left(-i\vec{\nabla}' + \frac{e}{\hbar c}\vec{A}(\vec{r}') + \frac{e}{\hbar c}\hat{\vec{a}}(\vec{r}')\right)^2 \hat{\Psi}_e(\vec{r}')\right] = \qquad (54)$$

$$= \phi^2 \frac{\hbar^2}{2m_e}\int d^2\vec{r}' \frac{\hat{\rho}_e(\vec{r}')}{|\vec{r}-\vec{r}'|^2}\hat{\Psi}_h(\vec{r}) - \frac{e}{c}\frac{d\hat{\omega}_e(\vec{r})}{dt}\hat{\Psi}_h(\vec{r}).$$

Formulas (53) and (54) constitute the basis for the main statements:

$$\left[\hat{\Psi}_e(\vec{r}), \hat{K}\right] = \frac{\hbar^2}{2m_e}\left(-i\vec{\nabla} + \frac{e}{\hbar c}\vec{A}(\vec{r}) + \frac{e}{\hbar c}\hat{\vec{a}}(\vec{r})\right)^2 \hat{\Psi}_e(\vec{r}) +$$

$$+\phi^2 \hat{L}(\vec{r})\hat{\Psi}_e(\vec{r}) + \frac{e}{c}\frac{d\hat{\omega}(\vec{r})}{dt}\hat{\Psi}_e(\vec{r}),$$

$$\left[\hat{\Psi}_h(\vec{r}), \hat{K}\right] = \frac{\hbar^2}{2m_h}\left(-i\vec{\nabla} - \frac{e}{\hbar c}\vec{A}(\vec{r}) - \frac{e}{\hbar c}\hat{\vec{a}}(\vec{r})\right)^2 \hat{\Psi}_h(\vec{r}) +$$

$$+\phi^2 \hat{L}(\vec{r})\hat{\Psi}_h(\vec{r}) - \frac{e}{c}\frac{d\hat{\omega}(\vec{r})}{dt}\hat{\Psi}_h(\vec{r}),$$

$$\hat{L}(\vec{r}) = \frac{\hbar^2}{2m_e}\int d^2\vec{r}' \frac{\hat{\rho}_e(\vec{r}')}{\left|\vec{r}-\vec{r}'\right|^2} + \frac{\hbar^2}{2m_h}\int d^2\vec{r}' \frac{\hat{\rho}_h(\vec{r}')}{\left|\vec{r}-\vec{r}'\right|^2},$$

$$\hat{\omega}(\vec{r}) = \hat{\omega}_e(\vec{r}) - \hat{\omega}_h(\vec{r}), \hat{\omega}_i(\vec{r}) = -\frac{\phi e}{\alpha}\int d^2\vec{r}' \theta(\vec{r}-\vec{r}')\hat{\rho}_i(\vec{r}').$$

(55)

The obtained results confirm the earlier derived equations of motion (39) for the dressed field operators.

### §4. The influence of the Chern-Simons gauge field on the energy level of the two-dimensional magnetoexciton

In the Landau gauge description, the two-dimensional electrons and holes are described as free moving particles along the in-plane $x$-axis, undergoing the Landau quantization along the in-plane $y$-axis, perpendicular to the x-axis. The free motion is represented by the plane wave functions with unidimensional (1D) wave vectors $p$ and $q$ as a quantum numbers, whereas the Landau quantization takes place in the form of harmonic oscillations around the gyration points situated on the y-axis at the distances $pl_0^2$ and $-ql_0^2$ from the origin, where $l_0$ is the magnetic length. The displacements of electrons and holes in the opposite parts of the y-axis are due to the different signs of their electric charges. The single particle wave functions corresponding to the lowest energy level of the Landau quantization is

$$\varphi_e(\vec{r}) = \frac{e^{ipx}}{\sqrt{L_x l_0 \sqrt{\pi}}}\exp\left[-\frac{(y-pl_0^2)^2}{2l_0^2}\right],$$

$$\varphi_h(\vec{r}) = \frac{e^{iqx}}{\sqrt{L_x l_0 \sqrt{\pi}}}\exp\left[-\frac{(y+ql_0^2)^2}{2l_0^2}\right],$$

(56)

$$S = L_x L_y.$$

Here we consider the 2D layer with the surface area $S = L_x L_y$. The wave functions (56) belong to the lowest levels of Landau quantization with quantum numbers $n_e = n_h = 0$, and we do not consider in this paper the excited Landau quantization energy levels. This approach is known as the lowest Landau Levels (LLL) approximation and the labels $n_e$ and $n_h$ at the electron field operators $a_p^+$, $a_p$ as well at the hole field operators $b_p^+$, $b_p$ will be dropped. The bare electron and hole field operators obey the Fermi statistics. The energies of the electrons and holes equal to $\frac{1}{2}\hbar\omega_{c,i}$, with $i = e, h$, where $\hbar\omega_{c,i}$ are the cyclotron frequencies. The electron and the hole field operators in the coordinate representation in the LLL-approximation are:

$$\hat{\Psi}_e(\vec{r}) = \frac{1}{\sqrt{L_x l_0 \sqrt{\pi}}} \sum_p e^{ipx} \exp\left[-\frac{(y - pl_0^2)^2}{2l_0^2}\right] a_p,$$

$$\hat{\Psi}_h(\vec{r}) = \frac{1}{\sqrt{L_x l_0 \sqrt{\pi}}} \sum_q e^{iqx} \exp\left[-\frac{(y + ql_0^2)^2}{2l_0^2}\right] b_q.$$
(57)

where we introduced the electron and the hole density operators

$$\hat{\rho}_e(\vec{r}) = \hat{\Psi}_e^+(\vec{r})\hat{\Psi}_e(\vec{r}) = \frac{1}{L_x l_0 \sqrt{\pi}} \sum_{p,q} e^{i(p-q)x} \exp\left[-\frac{(y - pl_0^2)^2}{2l_0^2} - \frac{(y - ql_0^2)^2}{2l_0^2}\right] a_q^+ a_p,$$

$$\hat{\rho}_h(\vec{r}) = \hat{\Psi}_h^+(\vec{r})\hat{\Psi}_h(\vec{r}) = \frac{1}{L_x l_0 \sqrt{\pi}} \sum_{u,v} e^{i(u-v)x} \exp\left[-\frac{(y + ul_0^2)^2}{2l_0^2} - \frac{(y + vl_0^2)^2}{2l_0^2}\right] b_v^+ b_u,$$
(58)

$$\hat{\rho}(\vec{r}) = \hat{\rho}_e(\vec{r}) - \hat{\rho}_h(\vec{r}).$$

Here the density operator of the electron-hole system $\hat{\rho}(\vec{r})$ was defined as the algebraic sum of the electron and hole density operators. This algebraic sum determines the C-S gauge field vector potential operator $\hat{\vec{a}}(\vec{r})$, as follows

$$\hat{\vec{a}}(\vec{r}) = -\frac{\phi e}{\alpha} \int d^2\vec{r}' \vec{\nabla}_{\vec{r}} \theta(\vec{r}-\vec{r}') \hat{\rho}(\vec{r}'),$$

$$\frac{e^2}{2m_e c^2} \hat{\vec{a}}^2(\vec{r}) = -\frac{\hbar^2 \phi^2}{2m_e} \int d^2\vec{r}' \int d^2\vec{r}'' \frac{(\vec{r}-\vec{r}')(\vec{r}-\vec{r}'')}{|\vec{r}-\vec{r}'|^2 |\vec{r}-\vec{r}''|^2} \hat{\rho}(\vec{r}') \hat{\rho}(\vec{r}''),$$

$$\theta(\vec{r}-\vec{r}') = \arctan\left(\frac{y-y'}{x-x'}\right);$$

$$\vec{\nabla}_{\vec{r}} = \vec{e}_x \frac{\partial}{\partial x} + \vec{e}_y \frac{\partial}{\partial x},$$

$$\vec{\nabla}_{\vec{r}} \theta(\vec{r}-\vec{r}') = \frac{-\vec{e}_x(y-y') + \vec{e}_y(x-x')}{|\vec{r}-\vec{r}'|^2},$$

$$\alpha = \frac{e^2}{\hbar c} = \frac{1}{137}.$$

(59)

To determine the influence of the C-S gauge field on the 2D magnetoexciton energy level the average value of the Hamiltonian (44) must be calculated, using the magnetoexciton wave function $|\Psi_{ex}(\vec{k})\rangle$, which was obtained in Ref [9]:

$$|\Psi_{ex}(\vec{k})\rangle = \Psi_{ex}^+(\vec{k})|0\rangle,$$

$$\Psi_{ex}^+(\vec{k}) = \frac{1}{\sqrt{N}} \sum_t e^{ik_y t l_0^2} a^+_{t+\frac{k_x}{2}} b^+_{-t+\frac{k_x}{2}},$$

(60)

$$N = \frac{S}{2\pi l_0^2}.$$

First, we will discuss the influence of the terms in the Hamiltonian (44), which are proportional to the square of the C-S vector potential $\hat{\vec{a}}(\vec{r})$ in the form

$$\frac{e^2}{2m_{e,h} c^2} \int d^2\vec{r}' \hat{\Psi}^+_{e,h}(\vec{r}) \hat{\vec{a}}^2(\vec{r}) \hat{\Psi}_{e,h}(\vec{r})$$

(61)

The average value of them calculated with the magnetoexciton wave function with wave vector $\vec{k}=0$ is:

$$\Delta_e = \langle \Psi_{ex}(0) | \frac{e}{2m_e c^2} \int d^2\vec{r} \hat{\Psi}_e^+(\vec{r}) \hat{\vec{a}}^2(\vec{r}) \hat{\Psi}_e(\vec{r}) | \Psi_{ex}(0) \rangle =$$

$$= \frac{\hbar^2 \phi^2}{m_e l_0^2} \cdot \frac{1}{8\pi^2} \int d^2\vec{\rho}_1 \int d^2\vec{\rho}_2 \frac{\vec{\rho}_1 \cdot \vec{\rho}_2}{\vec{\rho}_1^2 \cdot \vec{\rho}_2^2} \exp\left\{ -\frac{1}{4}\left( \vec{\rho}_1^2 + \vec{\rho}_2^2 + |\vec{\rho}_1 - \vec{\rho}_2|^2 \right) + \right.$$

$$\left. + \frac{i}{2}[\vec{\rho}_2 \times \vec{\rho}_1] \right\} = \frac{\hbar^2 \phi^2}{m_e l_0^2} \cdot \frac{1}{8\pi^2} \int_0^\infty d\rho_1 \int_0^\infty d\rho_2 \int_0^{2\pi} d\varphi_1 \int_0^{2\pi} d\varphi_2 \cos(\varphi_1 - \varphi_2) \times$$

$$\times \exp\left\{ -\frac{1}{2}\left[ \rho_1^2 + \rho_2^2 - \rho_1 \rho_2 \cos(\varphi_1 - \varphi_2) \right] + \frac{i}{2} \rho_2 \rho_1 \sin(\varphi_1 - \varphi_2) \right\}. \quad (62)$$

The Fourier series expansions of the exponents $\exp(iz\sin t)$ and $\exp(z\cos t)$ contain coefficients expressed through the Bessel functions $J_\nu(z)$ and the modified Bessel functions $I_\nu(z)$ [10]

$$e^{iz\sin t} = J_0(z) + 2\sum_{k=1}^{\infty} \left[ J_{2k}(z) \cos 2kt + i J_{2k-1}(z) \sin(2k-1)t \right],$$

$$e^{z\cos t} = I_0(z) + 2\sum_{k=1}^{\infty} I_k(z) \cos kt. \quad (63)$$

Substitution them in to the previous expression leads to its transformation

$$\Delta_e = \frac{\hbar^2 \phi^2}{m_e l_0^2} \left\{ \frac{1}{4\pi^2} \int_0^\infty d\rho_1 \int_0^\infty d\rho_2 \exp\left[ -\frac{1}{2}(\rho_1^2 + \rho_2^2) \right] J_0\left( \frac{1}{2} \rho_1 \cdot \rho_2 \right) \times \right.$$

$$\times I_1\left( \frac{1}{2} \rho_1 \cdot \rho_2 \right) + \frac{1}{2} \sum_{m=1}^{\infty} \int_0^\infty d\rho_1 \int_0^\infty d\rho_2 \exp\left[ -\frac{1}{2}(\rho_1^2 + \rho_2^2) \right] J_{2m}\left( \frac{1}{2} \rho_1 \cdot \rho_2 \right) \times \quad (64)$$

$$\left. \times \left[ I_{2m+1}\left( \frac{1}{2} \rho_1 \cdot \rho_2 \right) + I_{2m-1}\left( \frac{1}{2} \rho_1 \cdot \rho_2 \right) \right] \right\}.$$

In subsequent calculations, we will use integrals with two Bessel functions [10]

$$\int_0^\infty x^{\alpha-1} e^{-px^2} J_\mu(bx) I_\nu(cx) dx = \frac{b^\mu c^\nu (p)^{-\frac{(\alpha+\mu+\nu)}{2}}}{2^{\mu+\nu+1} \Gamma(\mu+1)} \times$$

$$\times \sum_{k=0}^{\infty} \frac{1}{k!} \Gamma\left[ \begin{array}{c} k + \frac{(\alpha+\mu+\nu)}{2} \\ \nu+k+1 \end{array} \right] \left( \frac{c}{2p} \right)^{2k} \cdot {}_2F_1\left( -k, -(\nu+k), \mu+1, -\frac{b^2}{c^2} \right). \quad (65)$$

Its applications give rise to the energy shift $\Delta_e$ of the magnetoexciton energy level

$$\Delta_e = \frac{\hbar^2 \phi^2}{m_e l_0^2} \left\{ \frac{1}{16\pi^2} \sum_{k=0}^{\infty} \frac{1}{2^k (k+1)} \times \right.$$

$$\times {}_2F_1\left(-k, -(k+1), 1, -1\right) + \frac{1}{2} \sum_{m=1}^{\infty} \sum_{k=0}^{\infty} \frac{(2m+k)!}{2^{4m+k+2} \Gamma(2m+1)} \times$$

$$\times \frac{1}{k!(k+2m+1)} \cdot {}_2F_1\left(-k, -(k+2m+1), 2m+1, -1\right) + \quad (66)$$

$$+ \frac{1}{2} \sum_{m=1}^{\infty} \sum_{k=0}^{\infty} \frac{(2m+k-1)!}{2^{4m+k} \Gamma(2m+1)} \frac{1}{k!} \times$$

$$\left. \times {}_2F_1\left(-k, -(k+2m-1), 2m+1, -1\right) \right\}.$$

Side by side with the terms (61) containing the square of the C-S vector potential $\hat{\vec{a}}^2(\vec{r})$, one may estimate the contribution of the mixed term proportional to the scalar product of two vector potentials $\hat{\vec{a}}(\vec{r})$ and $\vec{A}(\vec{r})$. It is expressed by the average value

$$\left\langle \Psi_{ex}(\vec{k}) \left| \frac{e^2}{m_e c^2} \int d^2\vec{r}' \hat{\Psi}_e^+(\vec{r}') \hat{\vec{a}}^2(\vec{r}') \vec{A}(\vec{r}') \hat{\Psi}_e(\vec{r}') \right| \Psi_{ex}(\vec{k}) \right\rangle =$$

$$= \frac{\hbar eB\phi}{4\pi^2 m_e c l_0^4 N} \cdot \int d^2\vec{r}' \int d^2\vec{r}'' \frac{y'(y'-y'')}{\left|\vec{r}'-\vec{r}''\right|^2} \exp\left[ -\frac{1}{2} \left| \frac{\vec{r}'-\vec{r}''}{l_0} \right|^2 - \quad (67) \right.$$

$$\left. -\frac{(\vec{k}l_0)^2}{2} + (y'-y'')k_x - (x'-x'')k_y \right].$$

The shift of the magnetoexciton energy level, $\Delta_e'$, in the point $\vec{k} = 0$ due this term, can be calculated exactly

$$\Delta_e' = \frac{\hbar eB\phi}{4m_e c}. \quad (68)$$

In the case of the electron effective mass $m_e$ equal to the free electron mass $m_0$, and in the magnetic field strength $B = 10\,\text{T}$, and $\phi = 1$, the shift of the magnetoexciton energy level in the point $\vec{k} = 0$ due to the influence of the C-S gauge field can be estimated as $\Delta_e' = \frac{1}{4}\,\text{meV}$.

## § 5. Conclusions.

The origin of the Chern-Simons gauge field, as well as of the quantum point vortices is related with such collective motion in the 2D system, when the main role is played the angles $\theta(\vec{r} - \vec{r}')$ created by the reference vectors $(\vec{r} - \vec{r}')$ with some selected axis. The reference vectors describe positions of the particles at the points $\vec{r}'$ with the density operator $\hat{\rho}(\vec{r}')$. The coherent summation of the angles weighted with the density operator gives rise to the phase operator $\hat{\omega}(\vec{r})$, whereas the gradients of the angles and their weighted summation gives rise to the vector potential $\hat{\vec{a}}(\vec{r})$ of the Chern-Simons gauge field.

The unitary transformation operators $\exp\left(\pm \frac{ie}{\hbar c} \hat{\omega}(\vec{r})\right)$ acting on the bare electron and hole field operators lead to the formation of the dressed field operators representing the composite particles with a number $\phi$ of the attached quantum point vortices. The dressed field operators obey the Fermi or the Bose statistics depending on the parity of the numbers $\phi$ of the attached vortices. The Hamiltonian describing the composite particles and their interactions through the Coulomb forces as well as under the influence of the C-S gauge field was deduced, and equations of motion for the dressed field operators were derived. The influence of the Chern-Simons gauge field on the energy levels of the 2D magnetoexcitons was estimated.